\pdfoutput=1

\documentclass[journal=jacsat,manuscript=article]{achemso}

\usepackage[version=3]{mhchem} 

\author{J. J. Viennot}
\email{jeremie.viennot@lpa.ens.fr}
\author{J. Palomo}
\author{T. Kontos}
\email{kontos@lpa.ens.fr}
\phone{+33 (0)1 44322518}
\fax{+33 (0)1 44323540}
\affiliation{Laboratoire Pierre Aigrain, Ecole Normale Sup\'erieure, CNRS UMR 8551, Laboratoire associ\'e aux universit\'es Pierre et Marie Curie et Denis Diderot, 24, rue Lhomond, 75231 Paris Cedex 05,
France}

\title[An \textsf{achemso} demo]
  {Stamping single wall nanotubes for circuit quantum electrodynamics}

\abbreviations{IR,NMR,UV}

\begin{document}


\begin{abstract}
 We report on a dry transfer technique for single wall carbon nanotube devices which allows to embed them in
 high finesse microwave cavity. We demonstrate the ground state charge readout and a quality factor of about $3000$ down to the single photon regime. This technique allows to make devices such as double quantum dots which could be instrumental for achieving the strong spin photon coupling. It can easily be extended to generic carbon nanotube based microwave devices.
 \end{abstract}

Understanding the dynamical properties of quantum coherent conductors is central in the field of nanoelectronics. In that context, carbon nanotubes are particularly interesting as model system for strongly correlated electrons and/or for encoding quantum information \cite{Delattre:09,Basset:11,Viennot:13}. In addition, they can be combined with various electrodes exhibiting electronic orders such as ferromagnetism or superconductivity \cite{Sahoo:05,Pillet:11}. They offer in principle a very powerful platform for probing the dynamics of mesoscopic systems emerging from hybrid structures.

One very promising avenue for probing the dynamics of mesoscopic systems is to embed them in a high finesse microwave cavity and ultimately use the tools of cavity quantum electrodynamics \cite{Delbecq:11,Frey:12,Petta:12,Toida:13}. The fabrication of such structures is a priori an experimental challenge since the superconducting microwave cavities made of metals such as Al or Nb are not compatible with the growth conditions of CVD carbon nanotubes (for example $900^{\circ}C$ with a flow of $CH_{4}$). In addition, the residues of growth of carbon nanotubes (for example catalyst or amorphous carbon) provide in general very strong dissipative media for microwave signals, resulting in poor microwave properties of the devices\cite{Delbecq:11,Delbecq:13}.

Here, we demonstrate a dry transfer (stamping) technique which allows us to solve these problems. Specifically, we are able to combine high finesse microwave cavities with high quality carbon nanotube based devices. We demonstrate the interest of our technique with a very closed double quantum dot with ferromagnetic contacts. Our technique has the advantage of simplicity over previous stamping techniques \cite{Pei:12}. It is also provides a high yield of single wall carbon nanotube transfer. It could easily be extended to other microwave setups with nanotubes \cite{Basset:11} or other type of nanowire.

\begin{figure}[!hpth]
\centering\includegraphics[height=0.35\linewidth,angle=0]{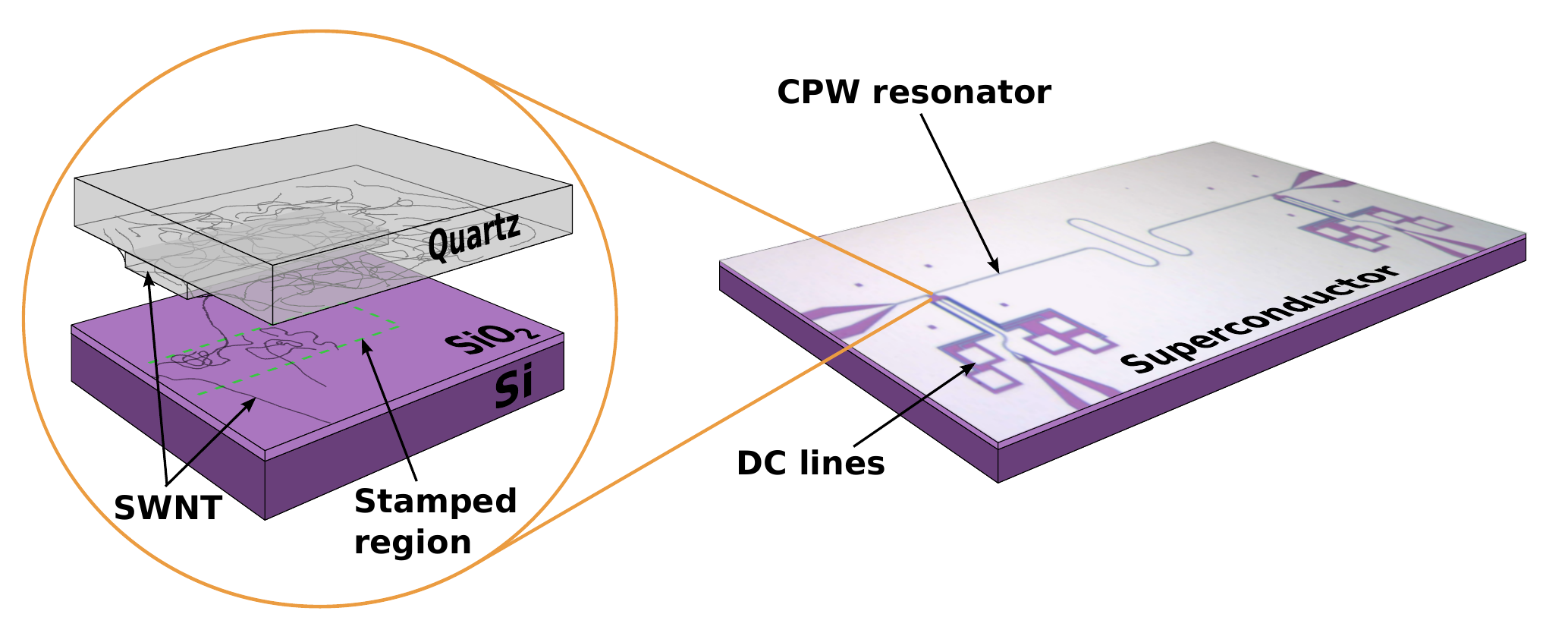}
\caption{ Principle of the stamping technique for combining high finesse microwave cavity with CVD-grown single wall carbon nanotubes (SWNT). The standart CVD growth is performed on a quartz substrate containing mesas which we use as stamps in order to transfert SWNT's on a clean silicon RF substrate. This can be done at chosen locations, in this case in  the ground plane openings of a superconducting co-planar microwave resonator.}%
\label{Figure1:Principle}%
\end{figure}

The principle of our stamping technique is depicted on figure \ref{Figure1:Principle}. The single wall nanotubes are grown using a methane process on a quartz substrate containing etched quartz islands. These islands are aligned with respect to Nb alignment markers made on a Si RF substrate and stamped \cite{Pei:12,ChineseNanolett:11} onto the Si substrate in locations matching the ground plane openings which one can see in figure 1a. Second, the nanotube device is fabricated using standard lithography techniques. The microwave cavity (made of Al or Nb) which is shown in figure 1 right panel can be fabricated after or before the fine structures. For the device described in details here, it was made of a 100 nm-thick Al and fabricated at the end of the whole nanofabrication process.

\begin{figure}[!hpth]
\centering\includegraphics[height=0.35\linewidth,angle=0]{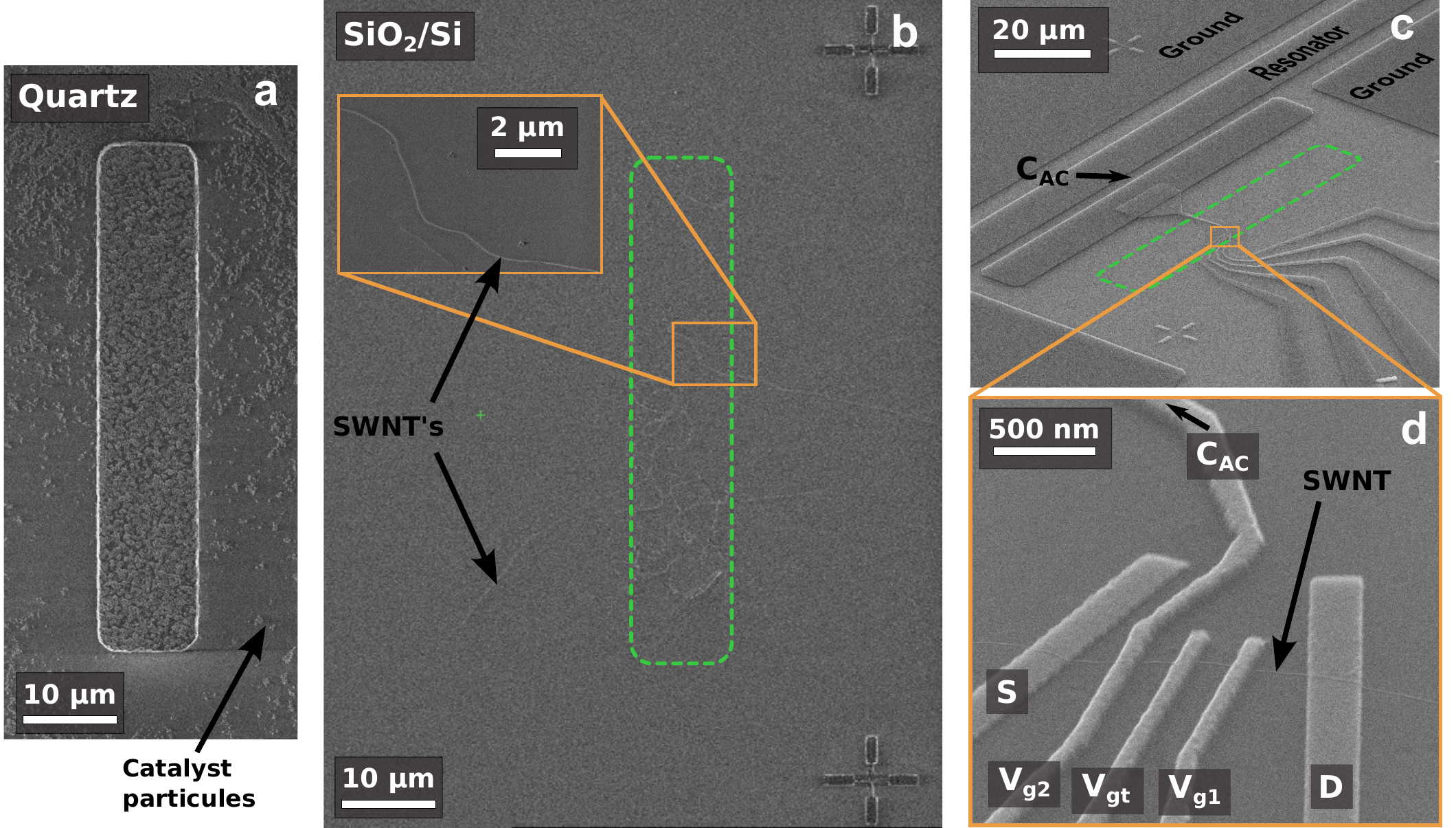}
\caption{  SEM images of the key fabrication steps: \textbf{(a)} mesa on the quartz substrate after catalyst deposition; \textbf{(b)} Region of the silicon subtrate where stamping has been performed. Green dashed rectangle is the print of the stamp. Inset: zoom on a SWNT used for fabricating the DQD device. \textbf{(c)} Tilted view of the device after e-beam lithographies and metal deposition corresponding the electrodes of the DQD as well as photolithography and metal deposition for the resonator. \textbf{(d)} Same tilt as \textbf{(c)}, zoomed on the DQD.}%
\label{Figure2:SEM}%
\end{figure}

We now describe in details the stamping process. We first make the quartz stamps which have typical size of $100 \mu m \times 10 \mu m $ with $3 \mu m$ height. They are the simplest stamps that one can make and optimization of the process has led us to use variants of this geometry e.g. 8 stamps of $10 \mu m \times 10 \mu m \times 3 \mu m$ forming a square array with a $10 \mu m $ pitch. The quartz substrate is covered by two layers of 500 nm of PMMA and a thin layer of 10 nm of Al (to avoid charging effects). The stamps are then exposed using ebeam lithography together with alignment markers. After development, a layer of 400nm of Al is evaporated at $\approx 10^{-6} mbar$. After lift-off, the quartz substrate with resulting Al mask is etched with reactive ion etching (SF6) for overall 40 min. The remaining Al is then dissolved in a solution of $KOH$ of $pH \approx $, yielding mesa structures of  $3 \mu m$ height. Note that, at the same time as the stamp is etched, alignment markers are also formed in the etching process. They are used at a later stage to align the stamp to the desired location of RF Si substrate. The latter is prepared separately with sputtered Nb alignment markers matching exactly those of the quartz structure.

 The chemical vapor deposition (CVD) growth is carried out using a methane process. The catalyst is made using $39mg$ of $Fe(NO_3)_3-H_2 O$, $7.9mg$ of
$MoO_2$ and $32mg$ of $Al_2O_3$ nanoparticles diluted in $30mL$ of IPA. It is sonicated for 1h and sedimented for 45 min before spreading it on the quartz substrate. This procedure and catalyst composition yields a high density of catalyst islands on the quartz stamp after catalyst deposition, as can be seen on figure \ref{Figure2:SEM}a. This means that the density of single wall carbon nanotubes grown on the quartz stamp is rather high. After stamping the quartz on the $Si/SiO_2$ substrate (which can be any substrate in principle), several single wall nanotubes are transferred as can be seen in figure \ref{Figure2:SEM}b. Depending on the stamping process, the stamped region highlighted by the green dashed lines can contain a more or less dense arrangement of nanotubes. In the situation shown in figure \ref{Figure2:SEM}b, rather long SWNTs ($\approx 10 \mu m$) can be isolated within that region. Usually, SWNTs can also be found apart from the stamp region as shown in figure \ref{Figure2:SEM}b. They probably arise from the breaking of SWNTs grown between the bottom and the top of the quartz mesa. Once the stamp procedure is done, standard lithography techniques can be used to make the fine structures. An example is shown in figure \ref{Figure2:SEM}c and d. A double quantum dot structure with $Al/Al_2 O_3$ top gates and two non-collinear $Pd_{75}Ni_{25}$ ferromagnetic leads has been embedded in an Al microwave cavity. 

\begin{figure}[!hpth]
\centering\includegraphics[height=0.45\linewidth,angle=0]{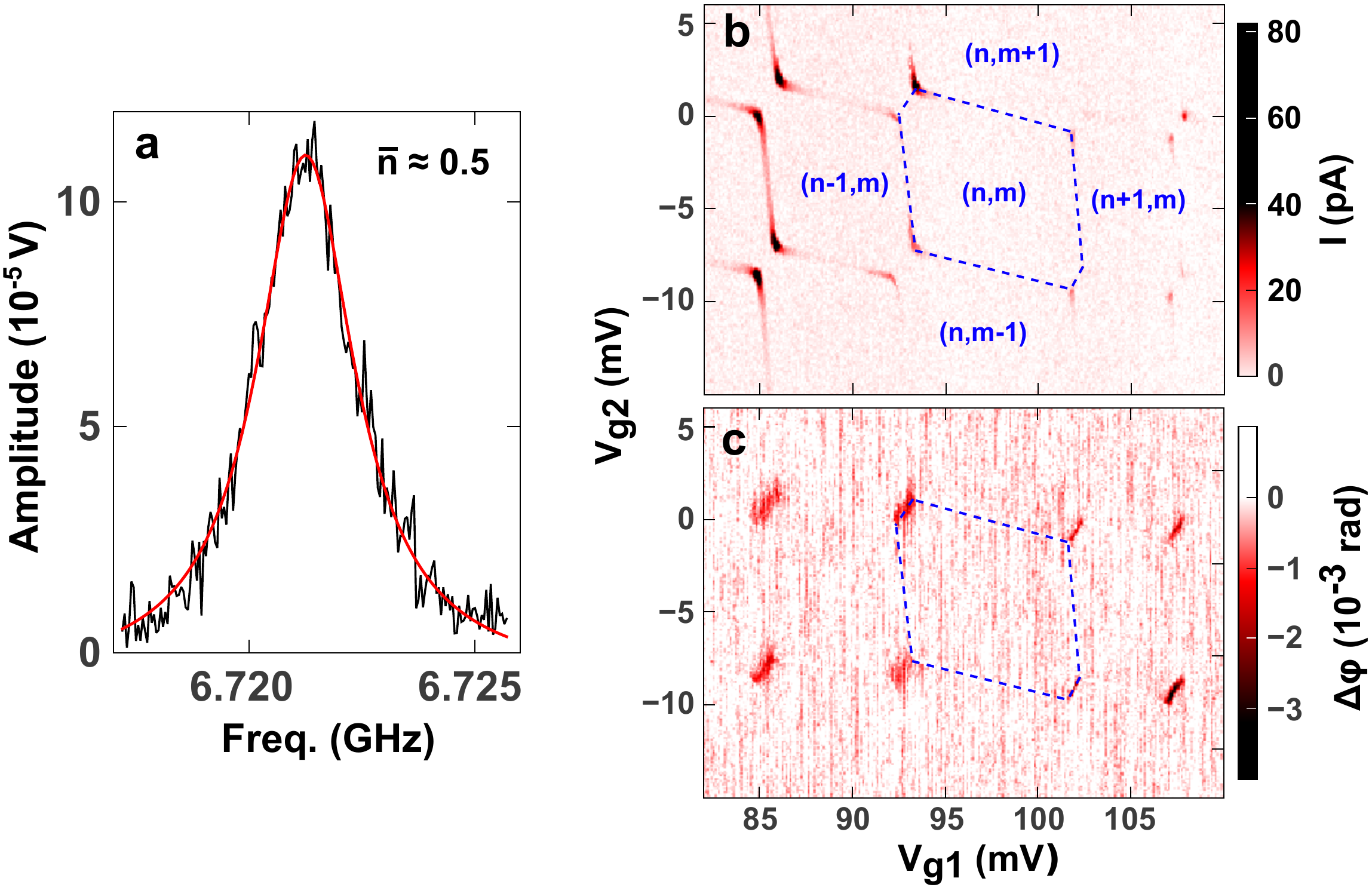}
\caption{ \textbf{(a)} Amplitude of the microwave field transmitted through the resonator in the single photon regime. \textbf{(b)} Direct current flowing trough the device measured at $V_{sd}=200 \mu V$ as a function of $V_{g1}$ and $V_{g2}$ ($V_{gt}=0mV$). The dashed lines follow co-tunneling lines, outlining the stability diagram of charge states (n, m) in the DQD. \textbf{(c)} Phase variation of the transmitted microwave signal measured at cavity resonance ($\simeq 6.72GHz$), simultanously with the direct current of \textbf{(b)}. }%
\label{Figure3:Cavity-DQD}%
\end{figure}

The resonance of the microwave cavity is shown in figure \ref{Figure3:Cavity-DQD}a. The measurement is carried out at $T=20mK$ (electronic temperature of about $40mK$). The cavity fundamental mode has a frequency $\omega_0/2\pi$ of about $6.72 GHz$ with a linewidth of $2.5MHz$, yielding a quality factor of about $3000$. Therefore, the cavity at thermal equilibrium is in the quantum regime ($\hbar \omega_0 \gg k_B T$). The joint measurement of the DC current and the phase of the microwave signal transmitted through the cavity at its resonance frequency are displayed in figure \ref{Figure3:Cavity-DQD}b top and bottom panel respectively. The DC current displays the usual honeycomb pattern of a double quantum dot (the device shown here is in the many electron regime). The phase of the microwave signal demonstrates the coupling of the cavity photons with the interdot transitions (internal to the double dot system). We therefore provide here another example using nanotubes, besides 2DEG\cite{Frey:12,Toida:13} and InAs nanowires \cite{Petta:12}, of coupling of closed double quantum dots to cavity photons. The depth of the tilted ticks allows to determine an effective spin/photon coupling strength of about $10 MHz$. The effective spin $\sigma_z$ is encoded in the hybridized bonding/antibonding states of the double quantum dot.

In order to use such an architecture in a circuit QED context, it is essential to be able to place the cavity in the single photon regime, while keeping its high finesse. Note that its is not trivial a priori since the losses of such coplanar waveguide cavities are known to be dominated at low input power by defects in the substrate. It is important in our case to establish that the stamping procedure does not add any further defect which might degrade more the quality factor of the cavity than in the high power (more than 1000 photon limit). The measurement presented here is done at an average coherent state photon number $\bar{n}$ of $0.5$. The photon number is calibrated using the power dependence of the phase contrast in the resonant regime \cite{Viennot:13}. The quality factor is 2500, which is only $30\%$ lower than the high power value of 3500. Note finally that our double quantum dot is connected with non-collinear ferromagnetic electrodes. This fact together with the possibility to contact nanotubes based double quantum dots with superconducting electrodes\cite{Herrmann:10,Schindele:12}, makes our setup ideal in order to investigate various kinds of highly desirable cQED architectures\cite{Cottet:10,Cottet:12}. Finally, our technique could be used in many other microwave setups using nanotube based devices\cite{Chaste:09}.

\begin{acknowledgement}

We gratefully acknowledge support and discussions from M. Rosticher, L.E. Bruhat, M.R. Delbecq, M.C Dartiailh and A. Cottet. This work was financed by the EU-FP7 project SE2ND[271554] and the ERC Starting grant CirQys.

\end{acknowledgement}



\end{document}